# Automated Classification of Research Papers Toward Sustainable Development Goals: A Boolean Query-Based Computational Framework

Sahil Dewani[1] and Kiran Sharma[1,2*]

[1]*School of Engineering & Technology, BML Munjal University, Gurugram, India*

[2]*Center for Advanced Data & Computational Science, BML Munjal University, Gurugram, India*

*Correspondance\**: *kiran.sharma@bmu.edu.in*

**Abstract**

The rapid expansion of scholarly publications across diverse disciplines has made it increasingly difficult to systematically evaluate how research contributes to the United Nations Sustainable Development Goals (SDGs). Domain classification of research articles done manually through research experts is extremely impractical because of the number of publications, expensive in time and may not be consistent when done by human beings. This paper proposes an automated and rule-based computational model of classifying research papers based on SDGs with expert curated Boolean query mappings to overcome these challenges. The proposed system has a web-based interface to input data and display results, a backend application programming interface to do high throughput processing, and a Python-based classification engine which uses structured Boolean expressions to process bibliographic metadata (titles, abstracts, and keywords). The framework can be used to support single-paper-based classification and batch-based classification as well as offer clear and understandable outputs that clearly show what query parts motivated each SDG assignment. The experimental testing on massive bibliographic data sets has shown that the system can process thousands of research records in an hour with reproducible and consistent results. The proposed approach provides a viable solution to institutions, researchers and policymakers who are interested in analysis of research alignment with the goal of sustainability in a systematic fashion that would not involve the use of machine learning models whose inputs and outputs are not easily understandable.

# 1. Introduction

The United Nations 2030 Agenda is a world-wide program based on seventeen Sustainable Development Goals (SDGs) to help in solving both the social, economic, and environmental issues present in the countries, such as the eradication of poverty, enhancement of health among the populations, combating climate change and environmental sustainability in the long run [1]. Scientific research and innovation are at the forefront to attain these objectives since they provide evidence base to be used to make informed policy formulation, technological advancement, and transforming the society. In turn, the issue of academic research in the context of the SDGs has gained a significant importance across universities, funding organizations, policymakers, and international organizations.

Although there has been increased focus on sustainability-based studies, the systematic review of the SDG applicability of research articles has become a huge challenge. The amount of scholarly literature in the various fields has grown at a large great pace and manual sorting by the relevant experts in the field has become very impractical, time consuming, costly and subject to biased interpretation. Such constraints point out to the necessity of automated, scalable and transparent procedures that would allow mapping large research corpora to SDG categories in a consistent and repeatable way.

## 1.1 Mapping SDGs by Using Query-based and Rule-driven Approaches

One of the most widespread SDG classification strategies that have been adopted is rule-based or query-driven approaches. These approaches are based on preset Boolean query keywords based on official SDG targets, SDG indicators and policy documents [3]. The Aurora Universities Network was among the first institutional initiatives to develop structured SDG search queries designed to identify research outputs aligned with specific Sustainable Development Goals within bibliographic databases [4].

Similarly, Elsevier has released multiple iterations of SDG mapping query sets that are widely used in large-scale bibliometric analyses and global university ranking exercises [5]. The latest version, the Elsevier 2023 SDG mapping dataset is publicly accessible and offers both curated Boolean query definitions and large-scale SDG-labeled publication metadata of academic output for decades [9]. The above dataset, available at

https://elsevier.digitalcommonsdata.com/datasets/y2zyy9vwzy/1 has been used and became an important reference resource to reproduce SDG-related research analytics.

Thematic keywords are usually used with action-oriented terminology that is aligned with SDG targets in query-based methods. As an example, one can identify publications that are related to SDG 1 (No Poverty) with the help of such combinations as terms concerning poverty, inequality and social protection [6]. The key advantage of these methods is that they are transparent, and every SDG assignment is explicitly traced to the direct keywords in the text of the research. This interpretability has made them to be adopted in institutional research assessment and international benchmarking models [1].

## 1.2 Hybrid SDG Classification and Machine Learning-Based Methods

Alongside rule-based approaches, machine learning and natural language processing (NLP) techniques have also been explored for automated SDG classification. Supervised learning models trained on labeled datasets can capture semantic relationships beyond exact keyword matches, which can improve recall when research articles use diverse or implicit terminology [7]. Most of these models are based on SDG-labeled corpora created with the help of query-based systems, such as SDG mapping dataset of Elsevier [9], one of the most popular training and evaluation datasets.

More recent works have suggested ensemble and hybrid methods that combine the output of multiple classifiers or combine rule based and learning based methods [8]. *Text2sdg* shows the application of various SDG labeling systems (like Elsevier and Aurora mappings) to study alignment of research in parallel [7]. Although these methods are promising, they might necessitate many computational resources and large-scale parameter optimization.

One major drawback of machine learning based systems is that they are not interpretable. The decision taken on classification is usually not transparent on a case-by-case document basis, which restricts their applicability in audit-sensitive applications where SDG assignments need to be justified.

## 1.3 Comparative Studies and Revealed Research Gaps

The comparative studies have found that there is a significant difference in the SDG classification outcomes based on the selected methodology. Comparative studies of independent sets of queries and machine learning models have demonstrated that alternative methods can provide significantly different SDG distributions of the same research corpus [2]. These inconsistencies have brought out the lack of a universal norm of SDG mapping and have emphasized the significance of methodological candidness.

There are multiple sources that highlighted explainability, reproducibility, and feasibility of refinement as highly important necessities of SDG-related research evaluation tools. Although machine learning solutions can be more practical in certain situations, query-based systems are more accessible and auditable to an institution. Nonetheless, several current query-based systems are published in the form of static data sets or programs, which restrict their application among other stakeholders that are not technical.

**1.4 Motivation and Contribution**

These considerations have inspired this work by applying a clear, query-based approach to automated SDG classification. The suggested framework uses publicly available and expert-curated SDG query sets, such as the Elsevier 2023 SDG mappings [9] and combines them into a modular, web-based system of computation.

The main contributions of this work are the following ones:
1. To create a scalable pipeline to capture and process metadata of research using an interface that is easy to use and intuitive.
2. To deploy a Boolean classification engine that can execute many SDG-related subqueries on research texts.
3. To provide interactive visual analytics that allow users to explore SDG distributions and the associated classification confidence scores.

**2. Methodology**

**2.1   Overall System Architecture**

The architecture is based on three layers, namely presentation layer, integration layer, and classification engine. The flow of the data starts with the research metadata provided by a user and concludes with ranked SDG outputs and visual analytics (see figure 1).

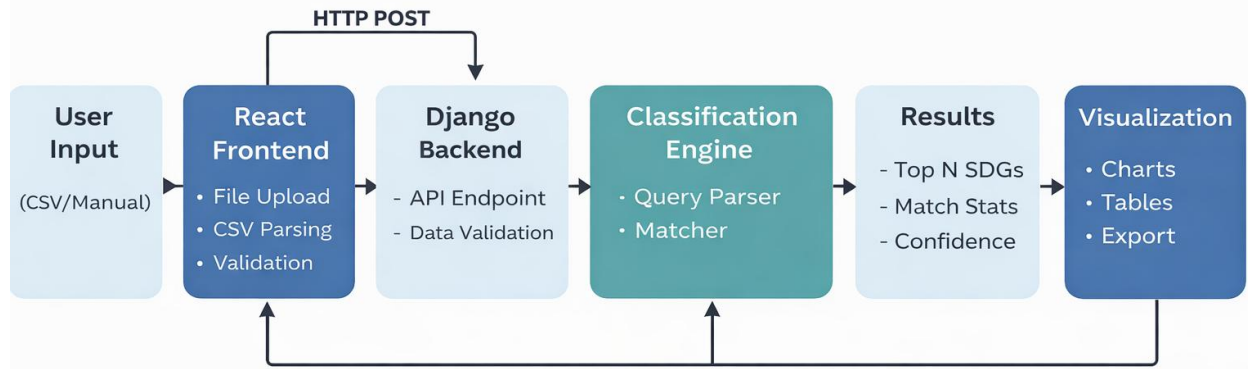

*Figure 1. Overall system architecture and data flow of the SDG classification framework*

### 2.1.1 Presentation Layer (Frontend Interface)

The presentation layer is a web-based frontend with React and modern UI libraries. It offers a convenient interface to the exploratory and large-scale SDG analysis. The frontend is backed by two major workflows:

- *Single-paper classification,* in which a user is required to manually fill in a paper title, abstract, and keywords.
- *Batch classification,* where users submit structured bibliographic files of multiple records (e.g. CSV or TSV).

The parameters that can be set by the user include the *number of top SDGs* to be returned (Top-N selection) and one can instantly see the results of classification. The findings are presented in ranked lists of SDG, confidence indicators, and interactive charts. Figure 2 & 3 shows the layout of the interface.

*Figure 2. Single paper classification: Web portal of the SDG classification.*

*Figure 3. Batch classification (Upload CSV): Web portal of the SDG Classification.*

### 2.1.2 Integration Layer (Backend API)

The integration layer is the interface that is between the classification engine and the user interface. It is deployed through a Python based RESTful API with Django. The backend API is to do validation of any incoming request, parsing of uploaded bibliographic files, normalizing fields in the metadata, calling the classification engine, and returning the structured SDG results in the form of a JSON.

#### (a) Data Ingestion and Preprocessing

The system helps in regular bibliographic formats of popular indexing databases such as Scopus and Web of Science. They are CSV and TSV files with the following mandatory fields:

- Paper title
- Abstract
- Author keywords
- Index keywords

Automatic column detection is performed using header-matching heuristics. For example, any column containing the term "abstract" (case-insensitive) is mapped to the abstract field. If automatic detection fails, users may manually map columns.

After ingestion, each paper record is consolidated into a single textual representation by concatenating all available metadata fields. Text preprocessing is deliberately lightweight:

- Conversion to lowercase
- Removal of punctuation (excluding Boolean operators)
- No global stopword removal
- No stemming or lemmatization

This design choice aligns with the structure of Boolean SDG queries, which rely on explicit keyword patterns and wildcard operators (e.g., sustainab*) rather than statistical language modeling.

#### (b) SDG Query Library Construction

The classification engine is built upon authoritative, expert-curated SDG query sets. The main source of this work was Elsevier SDG Mapping Dataset (2023), which is available to every person in the open access at the following source: https://elsevier.digitalcommonsdata.com/datasets/y2zyy9vwzy/1. It is a structured dataset in the form of Boolean queries that are in line with SDG targets and indicators and that are actively utilized in institutional research evaluation systems [9].

An SDG is represented by a few Boolean sub-queries, one to each thematic aspect of the goal. As an illustration, SDG 1 (No Poverty) has sub-queries that pertain to poverty eradication, income disparity, and social security.

Each sub-query may include:
- Logical operators (AND, OR)
- Wildcards
- Geographical limitations (where necessary)

### 2.1.3 Classification Engine

#### (a) Boolean Matching Logic

The classification engine then works in a loop, whereby each SDG is encountered, all the sub-queries related to it are checked against the consolidated paper text. A sub-query can be said to be matched when its Boolean conditions are met.

In the case of every SDG, the engine calculates:
- *Matched:* Number of Boolean sub-queries matched for a given SDG.
- *Total*: Total number of Boolean sub-queries defined for that SDG.

#### (b) SDG Scoring and Ranking

It is calculated using the ratio to obtain an SDG relevance score:

$$Score = Matched / Total$$

Every SDG is ranked according to this score. The system will default on displaying the top three SDGs of each paper, but this can be modified by the user. Most importantly, the detailed match statistics will also be returned by the system, and the users will be able to examine which sub-queries generated each assignment to SDGs.

#### (c) Visual Analytics

The analytics module converts the outputs of the classification into visual summaries that can be interpreted as shown in figure 4. It supports:
- SDG frequency bar charts, pie charts and donut charts.
- Tables of summarized classified papers.
- Reportable exportable result files.

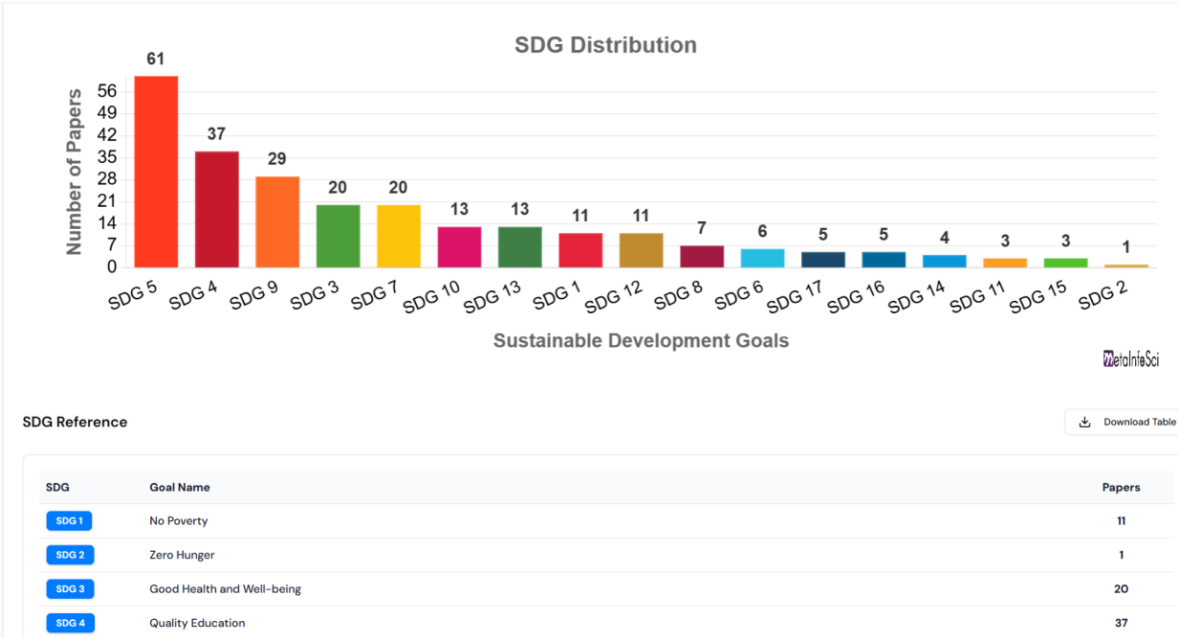

***Figure 4.*** *Visualization of SDG-wise distribution of classified papers, showing the frequency of each SDG within a sample corpus.*

## 2.2  System Overview and Design Principles

The proposed system is planned to be a highly customizable, query-based system to automatically classify research papers based on the United Nations Sustainable Development Goals. The following four principles form the central values of the methodology:

- *Transparency*:  All SDG assignments need to be resolved to clear keyword or Boolean rule matches.
- *Scalability*: The system should be capable of single-paper analysis as well as large scale batch processing.
- *Reproducibility*:  SDG output should always be the same when using the same inputs.
- *Extensibility*: Query definitions and classification logic must be maintainable without altering the underlying code.

To fulfill these needs, the framework is based on a layered web-based architecture, where user interaction, data processing and classification logic are separated. It is this separation that permits the Boolean classification engine to work without the user interface, permitting it to be reused in different analytical tasks.

## 2.3 Experimental Setup and Dataset Construction

Instead of testing the system on a large, noisy corpus, a controlled evaluation set was built, and the classification performance could be evaluated very accurately and in a manner that is understandable. Three Sustainable Development Goals have been chosen to be analyzed specifically:

- SDG 1: No Poverty
- SDG 4: Quality Education
- SDG 5: Gender Equality

On a case-by-case basis on the SDG, 50 articles were manually identified in the Scopus database. This left a total evaluation corpus of 150 research papers, where SDG relevance was known.

All papers were subjected to the SDG classification system in three slider settings:

- Slider = 1: Top-1 SDG returned
- Slider = 2: Top-2 SDGs returned
- Slider = 3: Top-3 SDGs returned

Such was a controlled design that allowed a methodical analysis of the effect of increasing the number of returned SDGs on the accuracy of classification and its coverage (see figure 5).

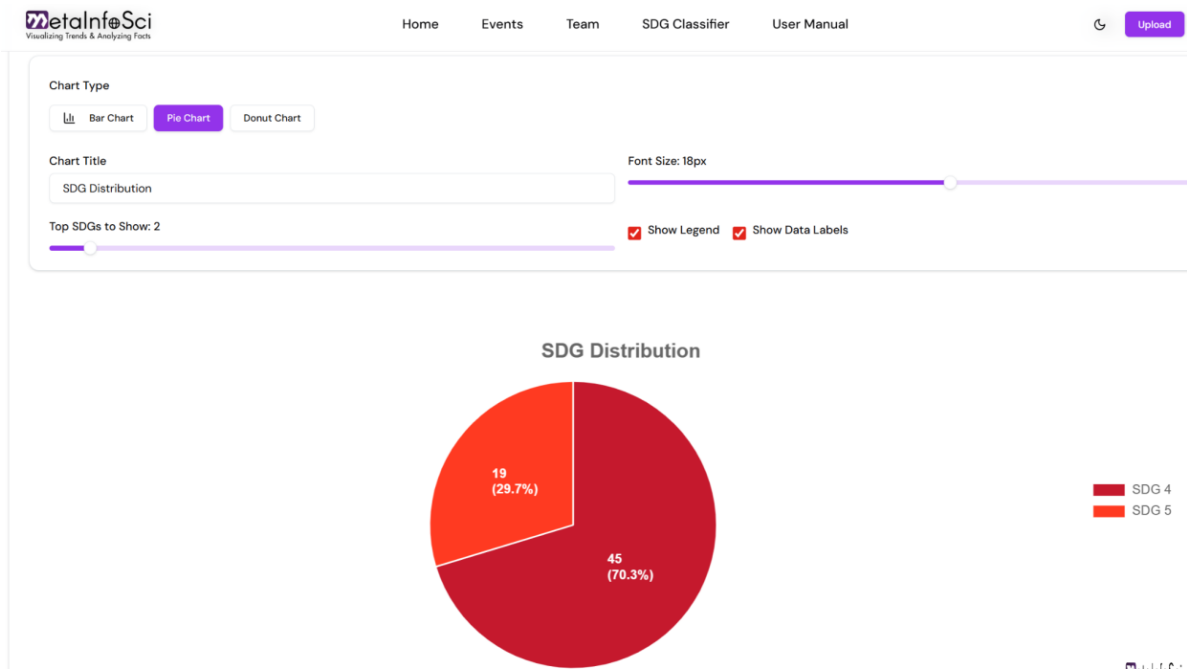

***Figure 5.*** *SDG distribution visualized using a pie chart with the slider set to Top-2 SDGs.*

## 3. Results and Discussion

This part is the empirical analysis of the offered SDG classification framework in the conditions of controlled experiments. The analysis is based on the behavior of classification depending on various slider settings (Top-1, Top-2 and Top-3 SDG outputs), quantitative congruence with SDG labels that are manually assigned, qualitative error tendencies, and the interpretability of the findings.

### 3.1 Selection of Top SDGs

**(a) Top-1 SDG Assignment**

The framework is conservative but accurate in the case when the slider is configured to give only the highest-ranked SDG. The returned SDG was in most instances directly associated with the manually assigned ground-truth label. Figure 6 shows the distribution of SDGs under the Top-1 set up. The distribution is heavily concentrated about the three SDGs assessed (SDG 1, SDG 4, and SDG 5) meaning that the Boolean scoring system was able to pinpoint the theme of sustainability that is prevailing in a given paper. This format focuses on the accuracy instead of the coverage thus it is specifically applicable when focusing on use cases that need one major SDG label to be used, like summary dashboards, institutional reporting, or high-level benchmarking.

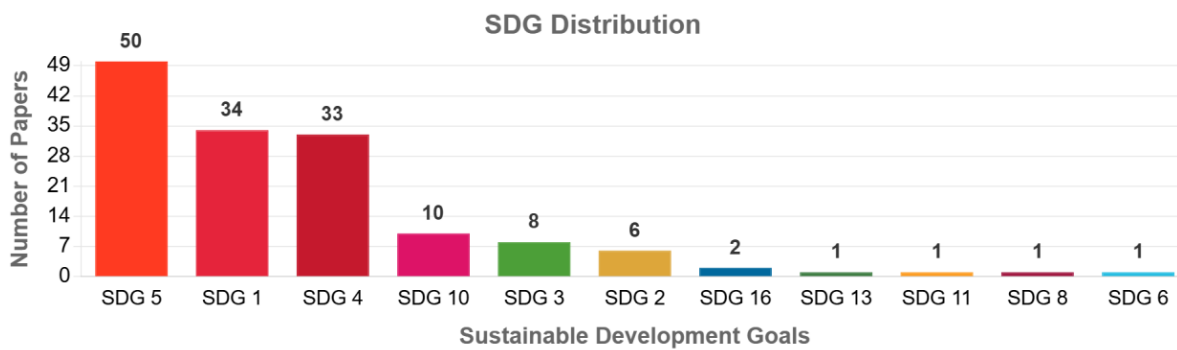

*Figure 6.* *Top-1 SDG per paper.*

**(b) Top-2 SDG Assignment**

The higher the slider value, the more relevant SDGs there are that can be displayed by the system per paper. This structure shows the multi-dimensionality of sustainability research, particularly in the case of papers that cut across several thematic sections. Figure 7 demonstrates that although the main SDGs (SDG 1, SDG 4, SDG 5) remain the most predominant ones, the secondary SDGs start getting more common. Education-oriented papers (SDG 4) tend to exhibit secondary congruency with SDG 5 regarding gender equity concerns, whereas poverty-oriented research (SDG 1) can have congruency with education or health-related objectives. This arrangement is a reasonable trade-off between recall and interpretability, hence it is an effective analysis tool in exploratory analysis and in self-assessing an institution.

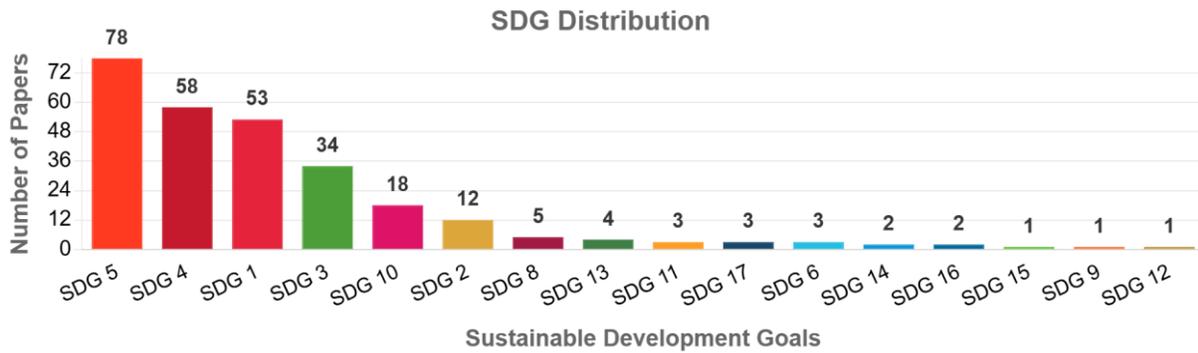

*Figure 7.* Top-2 SDGs per paper.

**(c) Top-3 SDG Assignment**

When the slider is adjusted to 3, the system will give as many SDGs as possible on a paper, which is 3, the most comprehensive picture of SDG relevance. This arrangement, as shown in Figure 8, has the added benefit of capturing numerous other contextual and secondary alignments, which are not predominant, but nonetheless, significant, as seen in sustainability research. Although overlaps between goals become more apparent, the framework's normalized scoring mechanism allows primary SDGs to remain clearly distinguishable from weaker or secondary associations. This framework is especially handy with respect to strategic analysis in which insight into secondary and tertiary sustainability interconnections can be of benefit.

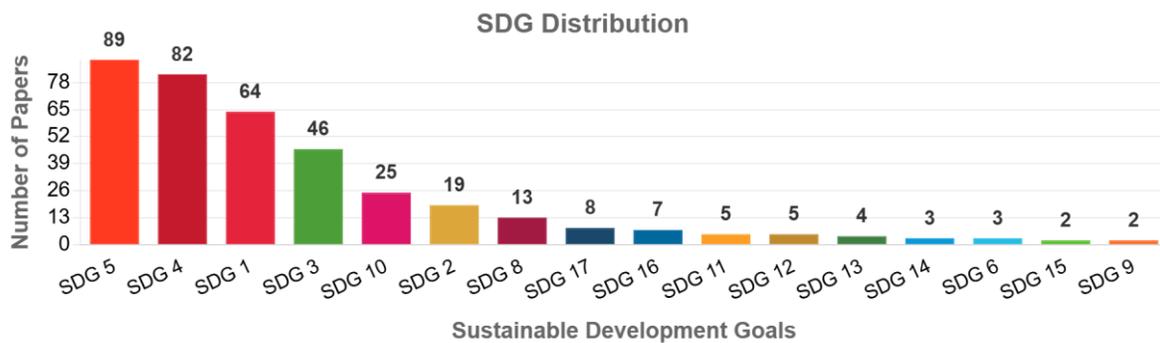

**Figure 8.** Top-3 SDGs per paper.

## 3.2 Quantitative Evaluation Metrics

To qualitatively assess classification performance, consistency between the manually assigned SDG ground truth and the system's output was evaluated under different slider settings. It was deemed that a prediction was correct at a given level of the slider when the ground-truth SDG was in the top-N returned SDGs.

*Table 1.* *The evaluation results for SDG 1, SDG 4, and SDG 5, including the number of correct classifications at each slider value and the number of papers for which no relevant SDG was recognized.*

| SDG | Total Papers | Correct @ Top-1 | Correct @ Top-2 | Correct @ Top-3 | Correct @ Top-4 | Correct @ Top-5 | No Recognition | Accuracy @ Top-1 (%) | Accuracy @ Top-2 (%) | Accuracy @ Top-5 (%) |
|---|---|---|---|---|---|---|---|---|---|---|
| SDG 1 | 50 | 32 | 46 | 50 | 50 | 50 | 0 | 64 | 92 | 100 |
| SDG 4 | 50 | 32 | 45 | 46 | 46 | 46 | 3 | 64 | 90 | 92 |
| SDG 5 | 50 | 39 | 49 | 50 | 50 | 50 | 0 | 78 | 98 | 100 |
| **Overall** | **150** | **103** | **140** | **146** | **146** | **146** | **3** | **68.67** | **93.33** | **97.33** |

## 3.3 Qualitative Analysis and Error Patterns

Qualitative review of misclassified and borderline cases shows that the major discrepancies are caused by overlap of the SDG scopes more than a false match of keywords. To illustrate, there are also education-focused articles that focus on social inclusion or gender equity, which results in SDG 5 secondary matches. Likewise, the themes of education or health can overlap with poverty-oriented research. Notably, the system gives complete visibility to such cases because it exposes the identical Boolean sub-queries and normalized scores to help the users comprehend why a certain SDG was proposed.

## 3.4 Interpretability and Practical Implications

A key strength of the proposed framework lies in its methodological clarity and consistency. The classification process is based on explicit Boolean query matching, where each SDG assignment follows predefined and expert-curated rules. Although end users do not directly view individual matched sub-queries through the interface, the underlying classification logic remains systematic, deterministic, and stable across repeated executions.

For each research paper, SDGs are ranked using the number of matched sub-queries and their normalized match scores. This rule-governed mechanism minimizes variability and supports reliable SDG identification across large research corpora.

## 4. Discussion and Conclusion

The present paper introduced a clear and generalizable computational framework of the automated classification of research articles based on the United Nations Sustainable Development Goals. Inspired by the accelerated development of academic literature and the recent rise of sustainability-related research evaluation, the suggested system deals with the main weaknesses related to manual classification and obscure automated solutions.

Using the Boolean query sets within the framework compiled by experts, specifically, the Elsevier SDG Mapping Dataset (2023) [9], provides the opportunity to make SDG assignments at scale, which can be reproduced and interpreted. The system architecture consists of a user-friendly web interface, an application programming interface with a robust backend application that can assess hundreds of SDG-related sub-queries in a document, and a dedicated Boolean classification engine.

Altogether, the experimental findings indicate that the suggested query-based framework can be an appropriate balance of precision, coverage, and transparency. Slider-based control allows the user to customize the system output to various analytical requirements, including conservative single-label assignment to a wider multi-SDG endeavor.

### 4.1 Key Findings and Observations

The SDG distribution observed in the analyzed dataset shows a clear concentration of research outputs in specific thematic areas. Studies connected to health, infrastructure, education, and climate measure seem to be the most common, meaning that these domains remain in the focus of scholarly researchers.

Second, the completeness and quality of input metadata have a strong impact on the results of classification. Articles containing abstracts and keywords are found to be categorized more confidently and covering the widest array of SDG than those that only contain titles. The above finding underscores the significance of detailed textual data in a sound automated SDG mapping.

Third, qualitative checking reveals that the results of the rule-based classification are close to the expert judgment in most situations. Notably, the framework is completely transparent because it gives the user a clear understanding of the specific query components that were used in each SDG assignment.

The proposed query-based framework offers clear practical advantages over machine learning-based SDG classification approaches. Rather than relying on opaque model behavior, the system follows a deterministic, rule-based process grounded in expert-defined Boolean queries. This ensures stable and consistent classification outcomes, which is particularly valuable in institutional research assessment, funding evaluation, and policy-oriented analysis where clarity and methodological reliability are essential.

### 4.2 Limitations

Although it has several strengths, there are some limitations with the proposed framework that can be mentioned. Currently, the system would only work on English-language research metadata. Consequently, the publications, which are published in other languages, might not be classified properly despite being highly oriented towards Sustainable Development Goals. This restriction is a result of the language contingency of the underlying Boolean query definitions, which are being maintained in the English language.

### 5. References

[1] Y. Kashnitsky, D. Meier, R. Mata, and D. U. Wulff, "Evaluating approaches to identifying research supporting the United Nations Sustainable Development Goals," *Quantitative Science Studies*, vol. 5, no. 2, pp. 408–427, 2024.

[2] A. C. S. Armitage, B. Jayabalasingham, R. M. K. N. H. Gunasekara, and E. Sugimoto, "Mapping scholarly publications related to the Sustainable Development Goals: Do independent bibliometric approaches get the same results?" *Quantitative Science Studies*, vol. 1, no. 3, pp. 1092–1108, 2020.

[3] D. U. Wulff, D. Meier, R. Mata, and Y. Kashnitsky, "Using novel data and ensemble models to improve automated labeling of Sustainable Development Goals," *Sustainability Science*, vol. 19, pp. 1773–1787, 2024.

[4] Aurora Universities Network, "Search Queries for Mapping Research Output to the Sustainable Development Goals (SDGs)," 2020. [Online]. Available: https://aurora-network-global.github.io/sdg-queries/

[5] B. Jayabalasingham, S. Ramanathan, and E. Sugimoto, "Identifying research supporting the United Nations Sustainable Development Goals," *Mendeley Data*, Version 1, 2019. doi:10.17632/87j5d7nh3b.1

[6] United Nations, "Transforming our world: The 2030 Agenda for Sustainable Development," United Nations General Assembly, New York, USA, 2015.

[7] D. Meier, R. Mata, D. U. Wulff, and Y. Kashnitsky, "text2sdg: An R package to monitor Sustainable Development Goals from text," *The R Journal*, vol. 17, pp. 50–70, 2025.

[8] D. Meier, D. U. Wulff, R. Mata, and Y. Kashnitsky, "Comparing and combining SDG classification systems for research evaluation," *Sustainability Science*, vol. 18, pp. 1459–1474, 2023.

[9] Elsevier Digital Commons, "Sustainable Development Goals (SDG) Mapping Dataset," 2023. doi:10.17632/y2zyy9vwzy.1 [Online]. Available: https://elsevier.digitalcommonsdata.com/datasets/y2zyy9vwzy/1